\documentclass[preprint,showpacs,preprintnumbers,amsmath,amssymb]{revtex4}

\usepackage{graphicx}
\usepackage{dcolumn}
\usepackage{bm}

\begin{document}

\preprint{}

\title{Eigenvalues, Separability and Absolute Separability of
Two-Qubit States}

\author{Paul B. Slater}%
\email{slater@kitp.ucsb.edu}
\affiliation{%
ISBER, University of California, Santa Barbara, CA 93106\\
}%
\date{\today}

\begin{abstract}
Substantial progress has recently been reported in the determination of the Hilbert-Schmidt 
(HS) separability probabilities 
for two-qubit and qubit-qutrit (real, complex and quaternionic) 
systems. An important theoretical concept employed has been that of a {\it separability function}. It appears that if one
could analogously 
obtain separability 
functions parameterized by the {\it eigenvalues} of the density matrices 
in question--rather than the {\it diagonal entries}, as originally 
used--comparable progress could be achieved in obtaining separability probabilities based on the broad, interesting
class of monotone metrics (the Bures, being its most prominent 
[minimal] member). 
Though large-scale numerical estimations of such 
eigenvalue-parameterized functions have been undertaken,
it seems desirable also to study them in lower-dimensional specialized scenarios in which they can be {\it exactly} obtained.
In this regard, we employ an Euler-angle 
parameterization of $SO(4)$ derived by S. Cacciatori (reported in an  
Appendix)--in the manner of the $SU(4)$-density matrix parameterization of 
Tilma, Byrd and Sudarshan. We are, thus, able to 
find simple exact separability (inverse-sine-like) 
functions for two real two-qubit (rebit) 
systems, both  having three free eigenvalues and
one free Euler angle. We also employ the 
important
Verstraete-Audenaert-de Moor bound to obtain exact 
HS probabilities that a generic two-qubit state
is {\it absolutely} separable (that is, can not be entangled 
by unitary transformations). In this regard, we make copious use of
trigonometric identities involving the tetrahedral dihedral
angle $\phi=\cos ^{-1}\left(\frac{1}{3}\right)$.

{\bf Mathematics Subject Classification (2000):} 81P05; 52A38; 15A90; 28A75
\end{abstract}

\pacs{Valid PACS 03.67.-a, 02.30.Cj, 02.40.Ky, 02.40.Ft}
\keywords{eigenvalues, diagonal entries, $SO(4)$, two qubits,
arc-sine function,
Hilbert-Schmidt metric, Bures metric, minimal monotone metric, 
separability functions, absolute separability,
separable volumes, 
separability probabilities, 
spectral separability conditions, tetrahedral dihedral angle}

\maketitle
\tableofcontents
\section{Introduction}
\.Zyczkowski, Horodecki, Sanpera and Lewenstein (ZHSL) \cite{ZHSL} 
were the first, it is clear, to pose the interesting question of determining the probability that a generic two-qubit or qubit-qutrit state is separable \cite{ZHSL}. The present author has further pursued this issue
\cite{slaterHall,slaterA,slaterC,slaterOptics,slaterJGP,slaterPRA,pbsCanosa,
slaterPRA2,slater833,slaterJGP2}. In particular, he has addressed it when the measure placed over the two-qubit or qubit-qutrit states is taken to be the {\it volume element} of certain metrics of interest that have been attached 
to such quantum systems (cf. \cite[sec. 5]{wkw1}). (This parallels the use
of the volume element [``Jeffreys' prior] of the Fisher information metric 
in classical Bayesian analyses \cite{kass,kwek}.)
 The (non-monotone \cite{ozawa}) Hilbert-Schmidt 
\cite{szHS} 
and (monotone \cite{petzsudar}) 
Bures \cite{szBures} can be considered as prototypical
examples of such quantum metrics \cite[chap. 14]{ingemarkarol}.

One useful (dimension-reducing) 
device that has recently been developed, in this regard, is the concept of a {\it separability function}. (The dimension-reduction stems from integrating 
over--that is, eliminating-- certain [off-diagonal] parameters.) 
The integral of the product of the separability function and the corresponding jacobian function yields
the desired separability {\it probability} \cite[eqs. (8), (9)]{slaterPRA2}. 
In the initial such studies
\cite{slaterPRA2,slater833},
the separability functions were taken to be functions of 
(ratios of products of) the {\it diagonal entries} of the
$4 \times 4$ or $6 \times 6$ density matrix ($\rho$).
In the later paper \cite[sec. III]{slaterJGP2}, it was 
also emphasized that it would be desirable to find analogous
separability functions parameterized, alternatively, 
in terms of the {\it eigenvalues} of $\rho$.
The motivation for this is quite straightforward. ``Diagonal-entry-parameterized separability functions'' (DSFs) 
have proved quite useful in studying Hilbert-Schmidt 
separability probabilities. However, the formulas \cite{szBures} 
for the large, interesting class of monotone metrics (Bures/minimal 
monotone, Kubo-Mori, Wigner-Yanase,...) are expressed in terms of the 
eigenvalues, and {\it not} the diagonal entries of $\rho$. 
Until such "eigenvalue-parameterized separability functions'' 
(ESFs) 
are obtained, it does not appear that as much progress can be achieved as has been reported for the Hilbert-Schmidt metric employing DSFs \cite{slaterPRA2,slater833,slaterJGP2}.
(Let us take note here--although we are not aware of
any immediate relevance 
for the problems at hand--of the Schur-Horn
Theorem, which asserts 
that the increasingly-ordered vector of diagonal entries of an 
Hermitian matrix {\it majorizes} the increasingly-ordered vector of
its eigenvalues \cite[chap. 4]{hornjohnson} (cf. \cite{nielsenvidal}).)
In particular, we would hope to be able 
to test the conjecture [numerically suggested]
that the two-qubit Bures separability probability is 
$\frac{1680 (\sqrt{2}-1)}{\pi^8}
\approx 0.0733389$ \cite{slaterJGP}.

In \cite{slaterJGP2} we also 
undertook large-scale 
numerical (quasi-Monte Carlo) analyses in order to estimate the ESF 
for the 15-dimensional convex set of
two-qubit states. We have continued this series of analyses (also now for the 9-dimensional real two-qubit states). However, at this stage we have not yet been able to discern the exact form such a (trivariate) function putatively takes.
In light of such conceptual challenges, it appears that one possibly effective strategy might be to find {\it exact} formulas for ESFs 
in lower-dimensional contexts, where the needed computations can, in fact, be realized. (This type of ``lower-dimensional''
strategy proved to be of substantial suggestive, intuition-enhancing  
value in the analyses of DSFs. One remarkable feature found was 
that the number of variables naively expected [that is, $n-1$, when $\rho$ is 
$n \times n$] to be needed in DSFs can be {\it reduced} 
by substituting new variables that are ratios of products of diagonal entries--so the {\it individual} independent diagonal entries need not be further utilized. One of our principal goals is to determine whether or not similar 
reductive structures are available in terms of ESFs.)

Tilma, Byrd and Sudarshan have devised an ($SU(4)$) 
Euler-angle parameterization of the (complex) two-qubit states
\cite{tbs}. To simplify our initial 
(lower-dimensional) analyses, we have 
corresponded with Sergio Cacciatori, 
who kindly developed a 
comparable ($SO(4)$) parameterization 
for the 9-dimensional convex set of two-qubit {\it real} density matrices. 
(This derivation is presented in Appendix I.)
In this $SO(4)$-density-matrix-parameterization, 
there are six Euler angles and three (independent) 
eigenvalues, $\lambda_1, \lambda_2, \lambda_3$, with 
$\lambda_{4}=1-\lambda_{1}-\lambda_{2}-\lambda_{3}$ (cf. \cite[eq. (9)]{batleplastino1}).
\section{Determination of ESFs} \label{secESF}
\subsection{Example 1}
To this point in time, we have only been able to obtain exact ESFs 
when no fewer than five of the six
Euler 
angles are held fixed. We now present our first 
such example, allowing the Euler angle 
$x_{1}$ to be the free one, 
and setting (using, for initial simplicity, the midpoints of the indicated 
variable ranges (\ref{ranges})) 
$x_{4}=x_{6}=\pi$ and $x_{2}=x_{3}=x_{5}=\frac{\pi}{2}$.
We will, thus, to begin with, be studying density matrices of the form,
\begin{equation}   \label{denmatrix}
\rho= \left(
\begin{array}{llll}
 \lambda _1 & 0 & 0 & 0 \\
 0 & \cos ^2\left(x_1\right) \lambda _2-\sin ^2\left(x_1\right)
   \left(\lambda _1+\lambda _2+\lambda _3-1\right) & -\frac{1}{2} \sin
   \left(2 x_1\right) \left(\lambda _1+2 \lambda _2+\lambda _3-1\right) &
   0 \\
 0 & -\frac{1}{2} \sin \left(2 x_1\right) \left(\lambda _1+2 \lambda
   _2+\lambda _3-1\right) & \sin ^2\left(x_1\right) \lambda _2-\cos
   ^2\left(x_1\right) \left(\lambda _1+\lambda _2+\lambda _3-1\right) & 0
   \\
 0 & 0 & 0 & \lambda _3
\end{array}
\right).
\end{equation}
By the Peres-Horodecki positive-partial-transpose 
(PPT) condition \cite{asher,michal},
$\rho$ is separable if and only if (an irrelevant nonnegative factor being
omitted from the PPT determinant)
\begin{equation} \label{PPT}
\lambda _1 \lambda _3-\frac{1}{4} \sin ^2(2 x_{1}) \left(\lambda _1+2
   \lambda _2+\lambda _3-1\right){}^2 \geq 0.
\end{equation}
Since we have set $x_{2}=x_{3}=x_{5}=\frac{\pi}{2}$, the Haar measure 
(\ref{invariant}) simply reduces to unity.
Integrating this measure 
over the interval $x_{1} \in [0, 2\pi]$, while 
enforcing the separability condition (\ref{PPT})--as well as 
requiring the 
facilitating eigenvalue-ordering
$\lambda_{1} \geq \lambda_{2} \geq \lambda_{3} \geq \lambda_{4}$--we obtain 
the desired $ESF(\lambda_1,\lambda_2,\lambda_3)$.

This ESF is equal to {\it unity} under the constraints (recall 
that we set $\lambda_4=1-\lambda_1 -\lambda_2-\lambda_3$)
\begin{equation} \label{inversetrig}
\lambda _1+2 \lambda _2\geq 1\land 2 \lambda _2+\lambda _3+2
   \sqrt{\lambda _1-2 \lambda _1 \lambda _2}\geq \lambda _1+1
\end{equation}
and
\begin{equation} \label{inversetrig2}
\lambda _1+2 \lambda _2=1\lor 2 \lambda _2+\lambda _3+2 \sqrt{\lambda
   _1-2 \lambda _1 \lambda _2}>\lambda _1+1.
\end{equation}
These constraints, thus, define the domain of eigenvalues for which
{\it all} the possible density matrices 
of the form (\ref{denmatrix})--independently of the 
particular value of $x_{1} \in [0,\pi]$--are separable.

\begin{figure}
\includegraphics{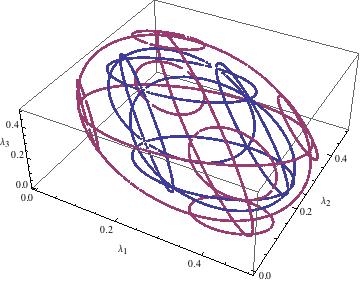}
\caption{\label{fig:VADpart}The blue points correspond to 
values of the (unsorted) eigenvalues
for which the VAD bound on absolute separability (\ref{VADbound})
is equivalent to the ZHSL purity bound,
$\lambda_{1}^2 +\lambda_{2}^2 +\lambda_{3}^2 +
\lambda_{4}^2  \leq \frac{1}{3}$. The red points are those 
which satisfy the VAD bound, and for which the purity is, then, the greatest possible, that is, $\frac{3}{8} > \frac{1}{3}$.}
\end{figure}

\begin{figure}
\includegraphics{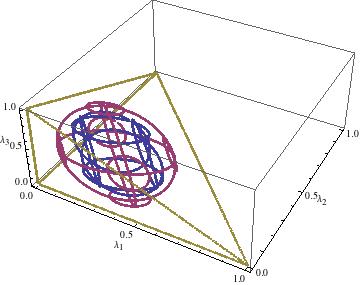}
\caption{\label{fig:simplex}Fig.~\ref{fig:VADpart} embedded in
the simplex of possible values of three (unordered) eigenvalues}
\end{figure}

Now, nontrivially, outside this domain of {\it total} separability 
(\ref{inversetrig}), (\ref{inversetrig2}), 
we have 
\begin{equation} \label{arcsin0}
ESF_{\{x_1\}}(\lambda_1,\lambda_2,\lambda_3)= 
\frac{2 \sin ^{-1} \tilde{U} }{\pi },
\end{equation}
where
\begin{equation}
\tilde{U} = \frac{2 \sqrt{\lambda _1 \lambda
   _3}}{\lambda _2-\lambda _4}
\end{equation}
and
\begin{equation} \label{constraints1}
\lambda _1+2 \lambda
   _2>1\land 2 \lambda _2+\lambda _3+2 \sqrt{\lambda _1-2 \lambda _1
   \lambda _2}<\lambda _1+1 .
\end{equation}
(The set of constraints (\ref{constraints1}), together with the 
imposed nonascending
order of the eigenvalues, ensure that the argument of the inverse sine
function, $\tilde{U}$,  is confined--sensibly--to the interval [0,1].)
The argument $\tilde{U}$ 
can be seen to be intriguingly analgous to the important 
(absolute) separability 
bound--originally suggested by Ishizaka and Hiroshima 
\cite{ishi}--of Verstraete, Audenaert and De Moor \cite{ver} 
\cite[eq. (3)]{roland},
\begin{equation} \label{VADbound}
VAD(\lambda_1,\lambda_2,\lambda_3)=
\lambda_{1}-\lambda_3 -2 \sqrt{\lambda_2 \lambda_4} <0,\hspace{.3in}
(\lambda_1>\lambda_2 >\lambda_3 >\lambda_4),
\end{equation}
or, equivalently,
\begin{equation} \label{otherform}
U \equiv \frac{2 \sqrt{\lambda_2 \lambda_4}}{\lambda_1-\lambda_3} > 1.
\end{equation}
(The associated 
greatest possible value of $\lambda_1$ is, then, $\frac{1}{2}$; of
$\lambda_2$, $\frac{2 +\sqrt{2}}{8}$; of $\lambda_3$, $\frac{1}{3}$; and of
$\lambda_4$, $\frac{1}{6}$.)
The inequality (\ref{VADbound}) improves upon
the ZHSL 
purity (inverse participation ratio) 
bound \cite{ZHSL} (Figs.~\ref{fig:VADpart} and \ref{fig:simplex}) 
(cf. (\cite{inclusion,clement,raggio})), 
\begin{equation}
\mbox{Tr}(\rho^2) =\lambda_{1}^2 +\lambda_{2}^2 +
\lambda_{3}^2 +\lambda_{4}^2  \leq \frac{1}{3}.
\end{equation}
\begin{figure}
\includegraphics{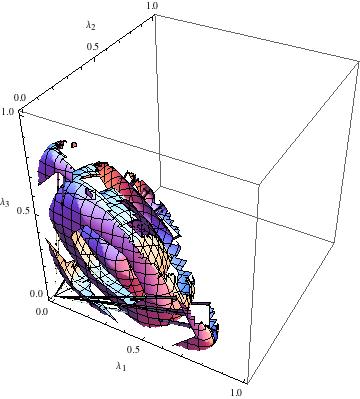}
\caption{\label{fig:contour}Three-dimensional contour plot of
$VAD(\lambda_1,\lambda_2,\lambda_3)$, given by 
eq. (\ref{VADbound}). Axes are of
length 1.}
\end{figure}
\begin{figure}
\includegraphics{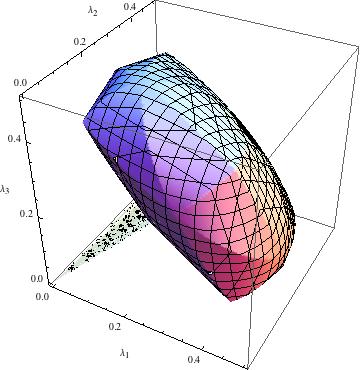}
\caption{\label{fig:contour2}The absolutely separable boundary 
surface $VAD(\lambda_1,\lambda_2,\lambda_3)=0$. Axes are of length $\frac{1}{2}$.}
\end{figure}
\begin{figure}
\includegraphics{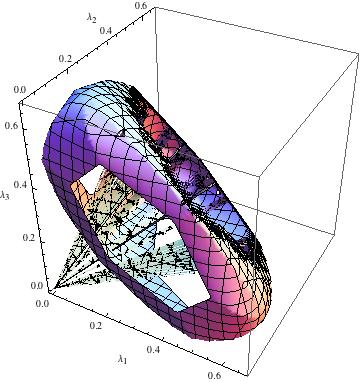}
\caption{\label{fig:contour3}The surface corresponding to 
non-absolutely separable states with
$VAD(\lambda_1,\lambda_2,\lambda_3)=\frac{1}{3}$. Axes are of length
$\frac{2}{3}$.}
\end{figure}
It appeared to us that the ESF for the {\it full} 9-dimensional
real and/or 15-dimensional complex convex sets of two-qubit states might
be of the form,
\begin{equation} \label{arcsin}
\frac{2 \sin ^{-1}\left(U\right)}{\pi},
\end{equation}
for $VAD(\lambda_1,\lambda_2,\lambda_3) \geq 0$, and simply unity
for $VAD(\lambda_1,\lambda_2,\lambda_3)<0$.
\subsection{Example 2}
For our second 
example with one free Euler angle and three free eigenvalues, 
we set $x_1=x_4=x_6=\pi$ and $x_2=x_3=\frac{\pi}{2}$, 
allowing $x_5$ to be the free Euler angle now. 
(Scenarios with only one of $x_2, x_3, x_4$ or $x_6$ allowed to be free
are trivially completely separable, and thus do not merit 
attention.) This results in density matrices of the form (again, taking $\lambda_4=1-\lambda_1-\lambda_2-\lambda_3$),
\begin{equation}
\left(
\begin{array}{llll}
 \lambda _3 \cos ^2(x_5)+\sin ^2(x_5) \lambda _1 & 0 & 0 &
   \cos (x_5) \sin (x_5) \lambda _3-\cos (x_5) \sin
   (x_5) \lambda _1 \\
 0 & \lambda _2 & 0 & 0 \\
 0 & 0 & -\lambda _1-\lambda _2-\lambda _3+1 & 0 \\
 \cos (x_5) \sin (x_5) \lambda _3-\cos (x_5) \sin
   (x_5) \lambda _1 & 0 & 0 & \lambda _1 \cos ^2(x_5)+\sin
   ^2(x_5) \lambda _3
\end{array}
\right).
\end{equation}
The Peres-Horodecki separability condition can now be expressed as
\begin{equation} \label{PH2}
-\frac{1}{4} \sin ^2\left(2 x_5\right) \left(\lambda _1-\lambda
   _3\right){}^2-\lambda _2 \left(\lambda _1+\lambda _2+\lambda
   _3-1\right) \geq 0.
\end{equation}
In contrast to our first example, however, the (reduced) 
Haar measure (\ref{invariant}) is now non-uniform,
being $\sin{x_5}$. We, thus, sought to integrate this measure over the 
range $x_5 \in [0,\pi]$, while enforcing the condition 
(\ref{PH2}), to obtain the corresponding ESF. 
However, Mathematica did not yield a solution after what we 
judged to be a reasonable amount of computer time to expend.

Therefore, we
made the transformation $x_{5} = \cos ^{-1}\left(-y_5\right)$,
which leads to a {\it uniform} measure for $y_5 \in [-1,1]$.
(There were some initial concerns expressed, in regard to this 
transformation, 
by  S. Cacciatori. He had written:
``The transformation you consider is not bijective on the whole
range of the parameters.
Indeed, the uniformization can be done only locally. This is because the
measure on a compact manifold cannot be an exact form (otherwise, using
Stokes Theorem, one finds zero volume for the manifold)'' 
(cf. \cite[p. 4394]{headrick}).
However, subsequent analysis revealed that nothing fallacious arises 
in this manner, since the results are equivalent to {\it twice} those
obtained by integrating over (the {\it half}-range) $y_5 \in [0,1]$.) 
Now, the region of total separability was of the form,
\begin{equation}
\lambda _1<\frac{1}{2}\land \lambda _1>0\land \left(\left(2 \lambda
   _1+\lambda _2>1\land (\text{M2}\lor \text{M3})\right)\lor
   (\text{M4}\land \text{M5})\right),
\end{equation}
where
\begin{equation}
M2=\lambda _1+2 \sqrt{\lambda _2-2 \lambda _1 \lambda _2}=2 \lambda
   _2+\lambda _3,M4=2 \lambda _2+\lambda _3+2 \sqrt{\lambda _2-2 \lambda _1 \lambda
   _2}>\lambda _1,
\end{equation}
\begin{equation}
M3=2 \lambda _2+\lambda _3+2 \sqrt{\lambda _2-2 \lambda _1 \lambda
   _2}>\lambda _1\land \lambda _1+2 \sqrt{\lambda _2-2 \lambda _1 \lambda
   _2}\geq 2 \lambda _2+\lambda _3
\end{equation}
and
\begin{equation}
M5=\left(2 \lambda _1+\lambda _2=1\land \lambda _1+\lambda _2+\lambda
   _3<1\right)\lor \left(\lambda _2>0\land \lambda _1+2 \sqrt{\lambda
   _2-2 \lambda _1 \lambda _2}>2 \lambda _2+\lambda _3\land 2 \lambda
   _1+\lambda _2<1\right).
\end{equation}
Also, non-trivially, the ESF took the (root) form 
\begin{equation} \label{root}
ESF_{\{x_5\}}(U) = \text{Root}\left[4 \text{$\#$1}^4-4
   \text{$\#$1}^2+U^2\&,3\right]-\text{Root}\left[4 \text{$\#$1}^4-4
   \text{$\#$1}^2+U^2\&,4\right]+1
\end{equation}
(Root[$f,k$] represents the k-{\it th} root of the polynomial 
equation $f[x]=0$) for
\begin{equation}
2 \lambda _1+\lambda _2>1\land 0<\lambda _1<\frac{1}{2}\land
   \lambda _1-2 \lambda _2-\lambda _3+2 \sqrt{\lambda _2-2 \lambda _1
   \lambda _2}<0\land \lambda _1+\lambda _2+\lambda _3<1 
\end{equation}
or 
\begin{equation}
   \frac{1}{2}<\lambda _1<1\land \lambda _1+\lambda _2+\lambda
   _3<1\land \lambda _2>0 .
\end{equation}
In fact, (\ref{root}) has an equivalent radical form, 
\begin{equation} \label{fallacious}
ESF_{\{x_5\}}(U) = \frac{\sqrt{1-\sqrt{1-U ^2}}}{\sqrt{2}}-\frac{\sqrt{1+\sqrt{1-U
   ^2}}}{\sqrt{2}}+1.
\end{equation}
The equivalence can be seen from a joint plot, 
in which (\ref{root}) amd (\ref{fallacious}) fully coincide. 
However, the Mathematica command ``ToRadicals''
did not produce (\ref{fallacious}), but gave the sign of the second of
the three addends as plus rather than minus. 
(``If Root objects in expr contain parameters, ToRadicals[expr] may yield a result that is not equal to expr for all values of the parameters''.)
This was clearly an
erroneous (``hyperseparable''--greater than 1) result. We initially 
had thought it might be due to an inappropriate uniformization, 
$x_{5} = \cos ^{-1}\left(-y_5\right)$. (In Appendix II, we also give a 
derivation of S. Cacciatori of (\ref{fallacious}) ``by hand''.) Therefore, 
we had also investigated alternative
approaches to obtaining the required ESF.

\begin{figure}
\includegraphics{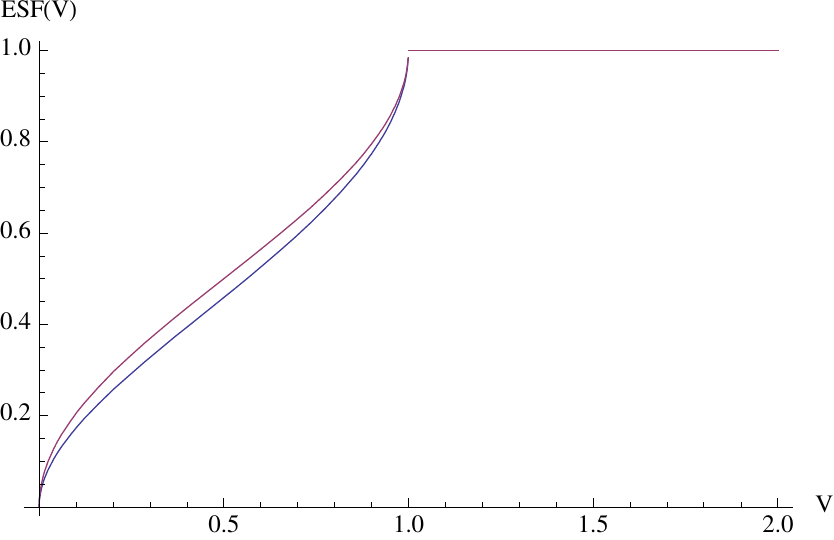}
\caption{\label{fig:ESF1}The eigenvalue-parameterized separability function
given by eq. (\ref{esfeq1})---along with the slightly subordinate 
function, (\ref{arcsin}), 
$ \frac{2 \sin ^{-1}\left(\sqrt{V}\right)}{\pi }, V \in [0,1]$---for 
the real two-qubit
scenario with the Euler angles $x_1=\pi, x_2 =\frac{\pi}{2},x_3=\frac{\pi}{2},
x_4=\pi,
x_6=\pi$ and $x_5$ free. Here, $V=U^2 = 
\frac{4 \lambda_2 \lambda_4}{(\lambda_1 -\lambda_3)^2}$, with
$\lambda_1 \geq \lambda_2 \geq \lambda_3 \geq \lambda_4$}
\end{figure}

We had reasoned that since the
transformed VAD variable $U \equiv \frac{2 \sqrt{\lambda_2 \lambda_4}}{\lambda_1-\lambda_3}$ 
seemed to play a vital role, we performed the 
change-of-variables,
\begin{equation}
\left\{\lambda _2\to \frac{1}{2} \left(-\lambda _1-\lambda
   _3+\sqrt{\left(\lambda _1+\lambda _3-1\right){}^2-V \left(\lambda
   _1-\lambda _3\right){}^2}+1\right)\right\},
\end{equation}
in our constrained integration analyses, where $V=U^2$.
Now, we again imposed the nonascending-ordering requirement on the eigenvalues 
(and their transformed equivalents), and the Peres-Horodecki condition 
(\ref{PH2}) 
(again discarding irrelevant nonnegative factors) became
\begin{equation} \label{better}
2 V+\cos \left(4 x_5\right)-1 \geq 0
\end{equation}
(conveniently being free of the individual $\lambda$'s).
Integrating the reduced Haar measure $\sin{x_5}$ 
over $x_5 \in
[0,\pi]$,  while enforcing (\ref{better}), 
we obtained the result (Fig.~\ref{fig:ESF1}),
\begin{equation} \label{esfeq1}
ESF_{\{x_5\}}(V) = \begin{cases}
 1 & V>1 \\
 -\sqrt{1-\frac{1}{\sqrt{2}}} \cos \left(\frac{1}{4} \sin ^{-1}(1-2
   V)\right)-\sqrt{1+\frac{1}{\sqrt{2}}} \sin \left(\frac{1}{4} \sin
   ^{-1}(1-2 V)\right)+1 & 0<V\leq 1.
\end{cases}
\end{equation}
We can represent this ESF more succinctly 
still, by using the variable
$W=1- 2 V$. Then, we have
\begin{equation}
ESF_{\{x_5\}}(W)=\begin{cases}
 1 & W<-1 \\
 2 \sin ^2\left(\frac{1}{8} \cos ^{-1}(W)\right)+\sin \left(\frac{1}{4}
   \cos ^{-1}(W)\right) & -1\leq W<1
\end{cases}.
\end{equation}
\section{Absolute Separability Analyses} \label{secAS}
\subsection{ZHSL uniform simplex measure}
Verstraete, Audenaert and De Moor remarked that the use of their result
(\ref{VADbound}) 
yields a better lower bound (assuming a uniform distribution on the simplex 
of eigenvalues) for the volume of separable states 
relative to the volume of all states: 0.3270 (as opposed to the 
ZHSL bound
of 0.3023 based on the purity \cite[eq. (35)]{ZHSL}) 
\cite[p. 6]{ver}. We are now able to give {\it exact} formulas for these
two bounds (certainly 
the second being new, and apparently somewhat challenging to 
derive), namely
\begin{equation}
ZHSL_{lower-bound}= \frac{\pi }{6 \sqrt{3}} \approx 0.30229989
\end{equation}
and (Figs.~\ref{fig:contour}, \ref{fig:contour2} and \ref{fig:contour2}) 
\begin{equation} \label{better2}
VAD_{lower-bound}= \frac{1}{4} \left(8-6 \sqrt{2}+3 \sqrt{2} \tan ^{-1}\left(\frac{1904
   \sqrt{2}}{5983}\right)\right) \approx 0.32723006.
\end{equation}

The Verstraete-Audenaert-de Moor bound (\ref{VADbound}) is based on 
the entanglement of formation. They also present another bound 
\cite[p. 2]{ver}
\begin{equation} \label{VADbound2}
\tilde{VAD}(\lambda_1,\lambda_2,\lambda_3) = 
\sqrt{(\lambda_1-\lambda_3)^2 +(\lambda_2-\lambda_4)^2} -\lambda_2-
\lambda_4<0, (\lambda_1>\lambda_2 >\lambda_3 >\lambda_4),
\end{equation}
based on the {\it negativity}. (There is also a more complicated
bound based on the {\it relative entropy of entanglement}. 
Since it involves logarithms, it is more difficult for us
to analyze.)
From this we obtained,
\begin{equation}  \label{more}
\tilde{VAD}_{lower-bound}= -\frac{12-8 \sqrt{2}+3 \pi +12 \cot
   ^{-1}\left(\frac{-6561+1904 \sqrt{2}}{5983}\right)}{4
   \sqrt{2}} \approx 0.32723006,
\end{equation}
{\it fully} equivalent, it would appear, to (\ref{better2}).
(Perhaps it is clear that these two bounds (\ref{VADbound}) and 
(\ref{VADbound2}) must be  equivalent, though no explicit mention is made of 
this, it seems in \cite{ver} nor the more recent \cite{roland}.
Mathematica readily 
confirmed for us that there is no nontrivial domain in which 
one of the bounds is greater than 0 and the other one, less than 0.)
Let us observe further, in regard to (\ref{more}) that $6561 =3^8$. 
Apparently,  there is a {\it right} triangle  with opposite legs of lengths
5983 and $1904 \sqrt{2}$ and hypothenuse $6561=3^8$, that is
of relevance in some way to
the issue of the absolute separability of two-qubit states.
We have consulted the Integer Superseeker website of N. J. A. Sloane 
(http://www.research.att.com/~njas/sequences) in regard to this issue, 
and found sequence A025172, entitled 
`` Let $\phi = \arccos{(\frac{1}{3})}$, the dihedral angle of the 
regular tetrahedron. Then $\cos{(n \phi)} = \frac{a(n)}{3^n}$''. 
In our application, $n=8$ and $a(n)=-5983$.
The accompanying comment is that the sequence is 
``Used when showing that the regular simplex is not `scisssors-dissectable' 
to a cube, thus answering Hilbert's third problem.''
Applying these facts to the problem at hand, we are able to obtain
the simplified results
\begin{equation}
VAD_{lower-bound}= \tilde{VAD}_{lower-bound} = 
\frac{1}{4} \left(8-6 \sqrt{2}+3 \sqrt{2} \left(\pi -8 \csc
   ^{-1}(3)\right)\right) \approx 0.32723006.
\end{equation}

We have obtained an {\it area-to-volume}
ratio of (cf. Fig.~\ref{fig:contour2})
\begin{equation}
R_{Area/Vol} = \frac{12 \left(-1+3 \sqrt{2} \cot ^{-1}\left(2
   \sqrt{2}\right)\right)}{8-6 \sqrt{2}+3 \sqrt{2} \left(\pi -8 \csc
   ^{-1}(3)\right)} 
= \frac{12 \left(-1+3 \sqrt{2} \cot ^{-1}\left(2
   \sqrt{2}\right)\right)}{8-6 \sqrt{2}+3 \sqrt{2} \pi -24 \sqrt{2} \cot
   ^{-1}\left(2 \sqrt{2}\right)} \approx 4.050415.
\end{equation}
\subsection{HS measure on two-qubit {\it real} density matrices}
Further, if we employ the Hilbert-Schmidt measure 
for two-{\it rebit}  density matrices \cite[eq. (7.5)]{szHS} 
on the simplex
of eigenvalues, we can obtain (V. Jovovic assisted with several 
trigonometric simplifications involving the dihedral angle, 
$\cos^{-1}(\frac{1}{3})$ an exact lower bound 
(much weaker than the conjectured actual value of $\frac{8}{17} 
\approx 0.470588$ \cite[sec. 9.1]{slater833}) on the Hilbert-Schmidt
two-qubit separability probability. This is 
\begin{equation} \label{HSlower}
VAD_{HS-lower-bound}^{real}= \frac{6928-2205 \pi }{16 \sqrt{2}} 
\approx 0.0348338.
\end{equation}
Also (cf. \cite{sbz}),
\begin{equation} \label{ratioReal}
R_{Area/Vol}^{HS-real}  = \frac{1}{1287 (-6928+2205 \pi )}
(\alpha_1+\alpha_2) \approx 12.489976122,
\end{equation}
where
\begin{equation*}
\alpha_1=34087768 \sqrt{2}-247867344 \pi -1292769261 \cos
   ^{-1}\left(\frac{1}{3}\right)
\end{equation*}
and
\begin{equation*}
\alpha_2=3226925088 \cot ^{-1}\left(\sqrt{2}\right)+3760128 \tan
   ^{-1}\left(\frac{1}{2 \sqrt{2}}\right)+350082810 \tan
   ^{-1}\left(\sqrt{2}\right).
\end{equation*}
The corresponding ratio based on the Bures metric is modestly larger 
(as will also be the case in the corresponding complex and 
quaternionic scenarios), 
that is, 14.582. Multiplying (\ref{ratioReal}) 
by the radius--$\frac{1}{12}$--of the maximal ball
inscribed inside the simplex of eigenvalues, we obtain a 
{\it dimensionless} ratio, $\gamma \approx 3.60555$ \cite[eq. (1)]{sbz} (cf. 
\cite{innami}).
\subsection{HS measure on two-qubit {\it complex} density matrices}
We have also obtained an exact expression for the HS absolute separability
probability of generic (complex) two-qubit states 
(using the indicated measure \cite[eq. (3.11)]{szHS}),
\begin{equation}
VAD_{HS-lower-bound}^{complex} = 
\frac{\psi_1+\psi_2}{32614907904} \approx 
0.0036582630543035
\end{equation}
where
\begin{equation*}
\psi_1=1959684729929728-1601255307608064 \sqrt{2}-1529087492782080 \sqrt{2} \pi
\end{equation*}
and
\begin{equation*}
\psi_2=45247615492565918250 \sqrt{2} \cot
   ^{-1}\left(\sqrt{2}\right)-22619730179635540245 \sqrt{2} \sec ^{-1}(3).
\end{equation*}
(The conjectured HS [absolute {\it and} non-absolute]
separability probability of generic two-qubit complex states
 is $\frac{8}{33} \approx 0.242424$ \cite[sec. 9.2]{slater833}.)
Further, we have
\begin{equation} \label{ComplexRatio}
R_{Area/Vol}^{HS-complex} 
= -3840 \frac{\tau_1+\tau_2}{\tau_3+\tau_4} \approx 20.9648519,
\end{equation}
with
\begin{equation*}
\tau_1=-5358569267936+33756573946095 \sqrt{2} \pi -270052591568760 \sqrt{2} \cot
   ^{-1}\left(\sqrt{2}\right),
\end{equation*}
\begin{equation*}
\tau_2=11149704525960 \sqrt{2} \cot ^{-1}\left(2 \sqrt{2}\right)+270052591568760
   \sqrt{2} \cot ^{-1}\left(3+\sqrt{2}\right),
\end{equation*}
\begin{equation*}
\tau_3=-1959684729929728+1601255307608064 \sqrt{2}+1529087492782080 \sqrt{2} \pi,
\end{equation*}
and
\begin{equation*}
\tau_4=-45247615492565918250 \sqrt{2} \cot
   ^{-1}\left(\sqrt{2}\right)+22619730179635540245 \sqrt{2} \sec ^{-1}(3).
\end{equation*}
The corresponding dimensionless ratio is $\gamma \approx 6.05203$,
while the Bures analogue of (\ref{ComplexRatio}) is 23.7826.

Employing the {\it Bures} measure \cite[eq. (3.18)]{szBures} for generic
two-qubit states
on the simplex of eigenvalues, we have, numerically, a lower bound--based 
on the absolutely separable states--on the two-qubit {\it Bures} 
separability probability of 0.000161792, considerably smaller 
than the conjectured value of $\frac{1680 (\sqrt{2}-1)}{\pi^8} 
\approx 0.0733389$ \cite{slaterJGP}.
\subsection{HS measure on two-qubit {\it quaternionic} density matrices}
The HS absolute separability probability of the
{\it quaternionic} two-qubit states is
\begin{equation} \label{QuatHS}
VAD_{HS-lower-bound}^{quat} = 
-\frac{13}{816946343106356485029888} \Sigma_{i=1}^{6} \zeta_i 
\approx 0.000039870347068,
\end{equation}
(the conjectured absolute {\it and} non-absolute 
separability probability being 
$ \frac{72442944}{936239725} 
\approx 0.0773765$ \cite[eq. (15)]{slaterJGP2}),
where
\begin{equation*}
\zeta_1=-216449750678398795533760757497856+176860737736399592490919645937664
   \sqrt{2},
\end{equation*}
\begin{equation*}
\zeta_2=279292548969739228073088142369304501839785 \sqrt{2} \pi
\end{equation*}
\begin{equation*}
\zeta_3=-558572941247617043110461841280869072896000 \sqrt{2} \cot
   ^{-1}\left(\sqrt{2}\right),
\end{equation*}
\begin{equation*}
\zeta_4=23637916932187025487103667523337320 \sqrt{2} \cot ^{-1}\left(2
   \sqrt{2}\right),
\end{equation*}
\begin{equation*}
\zeta_5= -16178155879591789043088455851252390200 \sqrt{2} \cot
   ^{-1}\left(3+\sqrt{2}\right)
\end{equation*}
\begin{equation*}
\zeta_6=-558589165778586158484606527963549721006600 \sqrt{2} \tan
   ^{-1}\left(\sqrt{2}\right).
\end{equation*}
Additionally,
\begin{equation} \label{QuatRatio}
R_{Area/Vol}^{HS-quat} = \frac{13}{3606947894919168 V^{quat}} 
(\psi_1+\psi_2) \approx 37.9283799507,
\end{equation}
where $V^{quat}$ is given by (\ref{QuatHS}) and
\begin{equation*}
\psi_1=-18147776040854148031593056-4720063928074960763823525 \sqrt{2} \pi,
\end{equation*}
and
\begin{equation*}
\psi_2=-37760511424599686110588200 \sqrt{2} \tan ^{-1}\left(9-7 
\sqrt{2}\right).
\end{equation*}
The corresponding dimensionless ratio is $\gamma \approx 10.948980$, 
and the Bures area-to-volume ratio counterpart to (\ref{QuatRatio}) 
is--slightly higher as in the real and complex cases--42.115.
\subsubsection{Two-qubit Cl{\'e}ment-Raggio spectral condition}
Cl{\'e}ment and Raggio have given a simple (linear) spectral 
condtion that ensures separability of two-qubit states 
\cite[Thm. 1]{clement},
\begin{equation}  \label{clement1}
3 \lambda_1 +\sqrt{2} \lambda_2+(3 -\sqrt{2}) \lambda_3 \leq 2, 
\hspace{.5in} \lambda_1 \geq \lambda_2 \geq \lambda_3.
\end{equation}
Based on this, we readily found--using the ZHSL uniform measure
on the simplex of eigenvalues--that a lower bound on the separability
probability is $\frac{1}{3 \sqrt{2}} \approx 0.235702$, and the
area-to-volume ratio is 6.

Now if we employed the HS measure for the real density matrices, 
along with the spectral condition (\ref{clement1}),
the lower bound is $\frac{104+75 \sqrt{2}}{17496} \approx 0.0120065$ 
and the area-to-volume ratio, 18. The corresponding two figures in
the HS complex measure case are $0.00060239769$ and 30, and, 
in the quaternionic case, $1.502473896 \cdot 10^{-6}$ and 54.

Cl{\'e}ment and Raggio also gave a spectral separability 
condition applicable to
any finite-dimensional system, but ``much weaker'' 
than (\ref{clement1}) in the specific
two-qubit case \cite[Thm. 2]{clement}. In the two-qubit case, it takes 
the form,
\begin{equation} \label{clement2}
3 \lambda_3 +3 \lambda_4 \geq 1, \hspace{.5in} \lambda_3 \geq \lambda_4.
\end{equation}
Based on this constraint, the lower bound on the separability probability using
the ZHSL uniform measure is $\frac{1}{9} \approx 0.11111$, while the 
area-to-volume ratio is 6. The associated separability probability based on
the HS measure on the real density matrices is $\frac{7}{6561} 
\approx 0.00106691$ and area-to-volume ratio, 18. 
The corresponding HS complex measure values are 
$\frac{143}{14348907} \approx 9.96592 \cdot 10^{-6}$ and 30, while the
two quaternionic values are $\frac{2185}{2541865828329} \approx 
8.59605 \cdot 10^{-10}$ and 54.

We see that the four area-to-volume ratios given are identical
using either of the two bounds (\ref{clement1}) and (\ref{clement2}), 
that is, 6, 18, 30 and 54..
\subsection{Qubit-{\it Qutrit} analyses}
\begin{figure}
\includegraphics{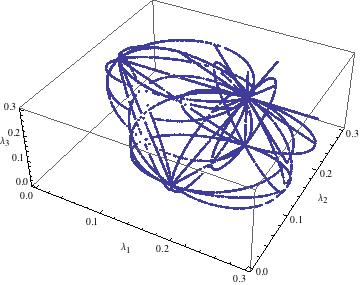}
\caption{\label{fig:contourQubQut}Three-dimensional projection of the sets
of six (unsorted) eigenvalues, $\lambda_i,=1,\ldots,6$, for which 
the Hildebrand bound (\ref{roland}) 
on absolute separability of qubit-qutrit systems is equivalent to
the ZHSL purity bound $\Sigma_{i}^{6} \lambda_i^{2} \leq \frac{1}{5}$ 
(cf. Figs.~\ref{fig:VADpart} and \ref{fig:simplex}).}
\end{figure}

The counterpart of the VAD bound (\ref{VADbound}) for qubit-{\it qutrit}
states was obtained by Hildebrand \cite[eq. (4)]{roland} 
(Fig.~\ref{fig:contourQubQut})
\begin{equation} \label{roland}
H(\lambda_1,\lambda_2,\lambda_3,\lambda_4,\lambda_5) = 
\lambda_1-\lambda_5 -2 \sqrt{\lambda_4 \lambda_6}<0, \hspace{.15in} 
(\lambda_1>\lambda_2 >\lambda_3 >\lambda_4 >\lambda_5>\lambda_6).
\end{equation}
We have also sought to obtain {\it exact} lower bounds on separability
probabilities here, using (\ref{roland}). 
Discouragingly, however, it seemed, after some
initial analyses that computer memory
demands may be too great to make any significant 
analytical progress on the {\it full} 
five-dimensional problem. 
(Nor have we been able to determine--as we have in the qubit-qubit case 
(after (\ref{otherform}))--the
maximum values that the $\lambda_i$'s can attain for absolutely separable 
states.)
However, we then sought to reduce/specialize 
the problem to more computationally
manageable forms.

Firstly, we found that if we set $\lambda_1=\frac{1}{3}$, then the 
absolute separability probability employing the uniform measure 
on the resulting four-dimensional simplex is 0.00976679. 
(The corresponding value for the full five-dimensional simplex
was reported by ZHSL as 0.056 \cite[eq. (35)]{ZHSL}.) 
However, this probability jumps dramatically to 0.733736, if we reduce
$\lambda_{1}$ from $\frac{1}{3}$ to $\frac{1}{4}$.
\subsubsection{Cl{\'e}ment-Raggio spectral conditions}
We did succeed, however, in this qubit-qutrit case in terms of the
``simple spectral condition'' of Cl{\'e}ment and Raggio 
\cite[Thm. 2]{clement}, which in our case takes the form,
\begin{equation}
3 \lambda_6 + 5 \lambda_5 \geq 1, \hspace{.5in} \lambda_5 \geq \lambda_6.
\end{equation}
The resultant bound on the ZHSL separability probability (based on the uniform 
measure over the simplex) is, then, 
quite elegantly, $\frac{1}{256}=2^{-8}$, and the associated
area-to-volume ratio is simply 15.
Additionally, we found that the lower bound for the separability
probability based on the
HS measure for real density matrices is
\begin{equation}
\frac{18989}{214748364800000} \approx 8.842442 \cdot 10^{-11}.
\end{equation}
Again, we have a simple area-to-volume ratio, 60.
\section{Discussion}
We have attempted, so far without success, to obtain exact ESFs with 
{\it more} than one free Euler angle. In particular, we have tried to
``combine'' the two {\it one}-free-Euler-angle scenarios analyzed above, by
letting {\it both} $x_1$ and $x_5$ be free.
The associated Peres-Horodecki separability criterion 
(cf. (\ref{PPT}), (\ref{PH2})) is then,
\begin{equation}
\left(\lambda _3 \cos ^2\left(x_5\right)+\sin ^2\left(x_5\right) \lambda
   _1\right) \left(\lambda _1 \cos ^2\left(x_5\right)+\sin
   ^2\left(x_5\right) \lambda _3\right)-\frac{1}{4} \sin ^2\left(2
   x_1\right) \left(\lambda _1+2 \lambda _2+\lambda _3-1\right){}^2 
\geq 0.
\end{equation} 
The best we were able to achieve, having first 
integrated over
$x_5 \in [0, \pi]$,
was a finding 
that the ESF for this {\it two}-Euler-angle scenario (again assuming 
$\lambda_{1} \geq \lambda_{2} \geq \lambda_{3} \geq \lambda_{4}$)
is expressible as
\begin{equation}
ESF_{\{x_1,x_5\}}(\lambda_1,\lambda_2,\lambda_3)= \int_{0}^{2 \pi}  
\begin{cases}
 1 & \kappa \geq 1 \\
 \cos \left(\frac{1}{4} \cos ^{-1}(\kappa )\right)-\sin \left(\frac{1}{4}
   \cos ^{-1}(\kappa )\right) & -1<\kappa <1
\end{cases}  d x_1
\end{equation}
where
\begin{equation}
\kappa= \frac{\cos \left(4 x_1\right) \left(\lambda _1+2 \lambda _2+\lambda
   _3-1\right){}^2+4 \lambda _2+2 \lambda _3-4 \lambda _2 \left(\lambda
   _2+\lambda _3\right)+\lambda _1 \left(-4 \lambda _2+4 \lambda
   _3+2\right)-1}{\left(\lambda _1-\lambda _3\right){}^2}.
\end{equation}
Alternatively, we were able to reduce this two-free-Euler-angle
problem to
\begin{equation}
ESF_{\{x_1,x_5\}}(\lambda_1,\lambda_2,\lambda_3)= 
\int_{0}^{\pi} \sin{x_5}  \begin{cases}
 1 & \eta >1 \\
 \frac{\sin ^{-1}(\eta )}{\pi }+\frac{1}{2} & -1<\eta \leq 1,
\end{cases} d x_5
\end{equation}
where
\begin{equation}
\eta= \frac{-\cos \left(4 x_5\right) \left(\lambda _1-\lambda _3\right){}^2+4
   \lambda _2+2 \lambda _3-4 \lambda _2 \left(\lambda _2+\lambda
   _3\right)+\lambda _1 \left(-4 \lambda _2+4 \lambda
   _3+2\right)-1}{\left(\lambda _1+2 \lambda _2+\lambda _3-1\right){}^2}.
\end{equation}

On the other hand, we have not yet found an 
exact ESF with even just one 
free Euler angle using the Tilma-Byrd-Sudarshan 
$SU(4)$ parameterization \cite{tbs} 
of the 15-dimensional convex set of
(in general, complex) $4 \times 4$ density matrices. 
Further, we have been able to convince ourselves--somewhat 
disappointingly--that 
the ESF for the real $4 \times 4$ density matrices can not simply be
a (univariate) function of $V$ (or $U$)--such as (\ref{arcsin}). 
We were able to reach this conclusion by finding distinct sets of
eigenvalues ($\lambda_1,\lambda_2,\lambda_3,\lambda_4$), which yielded
 the same identical value of $V$ (we used $V=\frac{1}{2}$),
but were capable of giving 
{\it opposite} signs for the determinant of the partial transpose, 
when sets of six (randomly-chosen) Euler angles were held fixed.

The analyses we have presented above have considerable similarities 
in purpose with a 
notable 
study of Batle, Casas and A. and A. R. Plastino, 
``On the entanglement properties of two-rebit systems'' \cite{batleplastino1} 
(cf. \cite{batlecasas1}). 
Some differences between our work and theirs are 
that we have  employed the
Verstraete-Audenaert-de Moor entanglement measure (\ref{VADbound}) and its transforms ($U$ and $V$) rather than the 
purity (which gives a smaller domain of absolute separability), 
and we have striven to obtain exact analytical results, 
 rather than
numerical ones (cf. \cite[sec. 3]{slaterJGP2}). 
(Let us note, in passing--although we have not adopted this viewpoint
here--``that there is qualitative difference between
the separability problems for real and complex matrices. In fact, the set
of separable density matrices has the same dimension as the full set of
density matrices in the complex case, but has lower dimension in the real 
case'' \cite[p. 713]{dahl}.)

We have been principally 
concerned here (sec.~\ref{secESF}) with the generation 
(by integration of the Haar measure over the Euler angles) of
{\it metric-independent} eigenvalue-parameterized separability functions 
(ESFs). Such ESFs could 
be applied, in conjunction with the formulas for arbitrary
metrics (e. g. \cite[eq. (4.1)]{szHS}, \cite[eq. (3.18)]{szBures} 
\cite[eqs. (14.34), (14.45)]{ingemarkarol}) 
to yield separability probabilities specific to the metrics in
question.

We have also reported (sec.~\ref{secAS}) 
advances in the study of absolute separability, 
implementing the Verstraete-Audenaert-de Moor bound (\ref{VADbound}).
A substantial analytical challenge in regard to this is the determination
of the volumes and bounding areas (and their ratios) of
the absolutely separable two-qubit states in terms of the Bures
(minimal monotone) metric. (We have given {\it numerical} values
of these ratios above.)
We have derived several remarkable exact trigonometric formulas pertaining
to absolute separability.
(V. Jovovic has assisted us in this task, simplifying 
Mathematica output by using identities involving the tetrahedral 
dihedral angle $\phi=\cos ^{-1}\left(\frac{1}{3}\right)$.) 
Nevertheless, we can certainly
not ensure that further (conceivably substantial) simplifications
exist.

\section{Appendix I (S. Cacciatori)}
The Lie algebra $so(4)$ of $SO(4)$ consists of the $4\times 4$ antisymmetric real matrices. A basis is given by the following matrices:
\begin{eqnarray*}
T_1=\left(
\begin{array}{cccc}
0 & 1 & 0 & 0\\
-1 & 0 & 0 & 0\\
0 & 0 & 0 & 0\\
0 & 0 & 0 & 0
\end{array}
\right)\\
T_2=\left(
\begin{array}{cccc}
0 & 0 & 1 & 0\\
0 & 0 & 0 & 0\\
-1 & 0 & 0 & 0\\
0 & 0 & 0 & 0
\end{array}
\right)\\
T_3=\left(
\begin{array}{cccc}
0 & 0 & 0 & 0\\
0 & 0 & 1 & 0\\
0 & -1 & 0 & 0\\
0 & 0 & 0 & 0
\end{array}
\right)\\
T_4=\left(
\begin{array}{cccc}
0 & 0 & 0 & 1\\
0 & 0 & 0 & 0\\
0 & 0 & 0 & 0\\
-1 & 0 & 0 & 0
\end{array}
\right)\\
T_5=\left(
\begin{array}{cccc}
0 & 0 & 0 & 0\\
0 & 0 & 0 & 1\\
0 & 0 & 0 & 0\\
0 & -1 & 0 & 0
\end{array}
\right)\\
T_6=\left(
\begin{array}{cccc}
0 & 0 & 0 & 0\\
0 & 0 & 0 & 0\\
0 & 0 & 0 & 1\\
0 & 0 & -1 & 0
\end{array}
\right)
\end{eqnarray*}.
The maximal proper subgroup of $SO(4)$ is $SO(3)$ and can be thought of as 
the isotropy subgroup of $SO(4)$. This means that if we consider $SO(4)$ as the group
of rotations of $\mathbb {R}^4$, then this group 
acts transitively on the unit sphere (translating the north pole everywhere on the sphere), but any point
is left fixed by the subgroup $SO(3)$. Thus we will have
$$
SO(4)/SO(3)\simeq S^3.
$$
This is the main observation needed to determine the range of the parameters. \\
Now let us look at the construction of the Euler parametrisation of the group.
First, one searches for a maximal subgroup, in our case $H=SO(3)$.
(In general, this might be not unique: for example for $E_6$ we have chosen $F_4$ as a maximal subgroup, but
another possible choice could have been $U(1)\times Spin (10)$). We can suppose that we 
know the Euler parametrisation for the subgroup. Otherwise, we can
proceed inductively, choosing a subgroup for $H$ and so on.\\
In our case for $H$, we can choose the $SO(3)$ subgroup generated by $T_i$, $i=1,2,3$. The Euler parametrisation of $H=SO(3)$ is well known,
$$
h[x,y,z]=e^{xT_3} e^{yT_2} e^{zT_3}\ .
$$
To determine the range of the parameters we first note that $SO(3)$ is not simply connected, but its universal covering is $SU(2)$ and
$SO(3)\simeq SU(2)/{\{I,-I\}}$ ($I$ being the identity). Also, it is well known that $SU(2)\simeq S^3$ as a manifold. With this in mind, let us compute
the invariant metric (and measure) on $H$. This can be done, noting that the Lie algebra is isomorphic to the tangent space at the identity $e$ of the
group. But our algebra is provided by a natural scalar product, the trace product
\begin{eqnarray*}
\langle\ , \rangle : so(4) \times so(4) \longrightarrow \mathbb{R},\\
\qquad (A,B) \longmapsto -\frac 12 {\rm Trace} (AB)\ .
\end{eqnarray*}
It is symmetric and positive definite, satisfies
$$
\langle T_I , T_J \rangle =\delta_{IJ}
$$
and induces a Euclidean scalar product on $T_e H$, with $\langle T_i , T_j \rangle =\delta_{ij}$. Thus, it is easy to compute the metric on $H$ as
the metric induced by the scalar product on $T_e H$. Indeed, if $h[x,y,z]$ is a generic point, then $dh$ is cotangent at the point $h$ and
$h^{-1} dh$ is cotangent at $e$, so that the metric tensor can be defined as
$$
ds_H^2 =\langle h^{-1} dh, h^{-1} dh \rangle =-\frac 12 {\rm Trace}( h^{-1} dh h^{-1} dh)\ .
$$
From this expression, we immediately see that $ds_H^2$ is both left and right invariant. A direct computation gives
$$
ds_H^2=dx^2+dy^2+dz^2+2\cos (y) dx dz .
$$
Now we need to compare this with the metric of a sphere $S^3$. This can be done, describing $S^3$ as the unit sphere in $\mathbb {C}^2$,
$$
S^3 =\{(z,w) \in \mathbb {C}^2| z\bar z +w \bar w =1 \}.
$$
Then, the point $p$ of $S^3$ can be parametrised as
$$
p=(\cos\lambda e^{i(\phi+\psi)}, \sin\lambda e^{i(\phi-\psi)})
$$
covered by $\lambda\in [0,\pi/2]$, $\phi \in [0,\pi]$, $\psi\in [0,2\pi]$. The Euclidean metric is
$$
ds^2 =dz d\bar z +dw d\bar w =d\lambda^2 +d\phi^2 +d\psi^2 +2 \cos (2\lambda) d\phi d\psi.
$$
Comparing this 
with the invariant metric, we see that it is the metric of $S^3$ with $y=2\lambda$, $x=2\phi$ and $z=2\psi$ (note that $x$ and $z$ can
be symmetrically interchanged). Now if we impose on $x,y,z$ the ranges for the sphere, that is
$$
x\in [0,2\pi], \qquad y\in [0,\pi], \qquad z\in [0,4\pi],
$$
we will cover $SO(3)$ exactly two times. This is because $SO(3)\simeq S^3/{\pm I}$ as noted before. However, we can easily check where duplication
takes place. Indeed,
$$
e^{zT3}=\left(
\begin{array}{cccc}
1 & 0 & 0 & 0\\
0 & \cos z & \sin z & 0\\
0 & -\sin z & \cos z & 0\\
0 & 0 & 0 & 1
\end{array}
\right)
$$
and takes all possible values twice when $z$ varies in $[0,4\pi]$. The correct range is then
$$
x\in [0,2\pi], \qquad y\in [0,\pi], \qquad z\in [0,2\pi].
$$
Now we are ready to construct the group $G=SO(4)$. Any element of $G$ can be written in the form
$$
e^{x T_4+y T_5 +z T_6} h[x_4,x_5,x_6].
$$
This is shown in an appendix of our paper on $F_4$ \cite{sergioF4}.
Here $(x,y,z)$ varies in $\mathbb {R}^3$. (We are not concerned at this level with the possibility of  covering the group many times. The only important point is that
this parametrisation is surjective). Now, we note that $H$ acts on $T_a$, 
as the rotations of $SO(3)$ act on the points of the unit sphere $S^2$.
In particular, from
$$
[T_2, T_4]=-T_6, \qquad [T_2, T_6]=T_4, \qquad [T_3, T_4]=0, \qquad [T_3, T_6]=T_5, \qquad [T_3, T_5]=-T_6
$$
we find
\begin{eqnarray*}
&& e^{xT_2} T_4 e^{-xT_2} =\cos x T_4 -\sin x T_6\ ,\\
&& e^{yT_3} e^{xT_2} T_4 e^{-xT_2} e^{-yT_3}=\cos x T_4-\sin x \cos y T_6 -\sin x \sin y T_5.
\end{eqnarray*}
Because any vector in $\mathbb {R}^3$ can be written as
$$
\vec x=r (\cos x, -\sin x \sin y, -\sin x, \cos y),
$$
and because 
$$
g a^A g^{-1}= e^{gAg^{-1}},
$$
we can write
$$
e^{x T_4+y T_5 +z T_6}=e^{x_1T_3} e^{x_2T_2} e^{x_3 T_4} e^{-x_2T_2} e^{-x_1T_3}
$$
for some $x_1, x_2$ and $x_3 (=r)$. Moreover, $e^{-x_2T_2} e^{-x_1T_3}\in H$, so that we can reabsorb it into $h$ and write the general point on $SO(4)$
as
$$
g[x_1,\ldots,x_6]=e^{x_1T_3} e^{x_2T_2} e^{x_3 T_4} h[x_4,x_5,x_6].
$$
We know the range for the parameters $x_4,x_5,x_6$. To determine the ranges for the remaining parameters, we note that
$$
p[x_1,x_2,x_3]=e^{x_1T_3} e^{x_2T_2} e^{x_3 T_4},
$$
parametrises the points of $SO(4)/SO(3)=S^3$, so that we need to compute the metric induced on $SO(4)/SO(3)$ and compare it with the metric of $S^3$. As before, we need to compute the left invariant form
$$
J_P:= P^{-1} dP.
$$
However, in general, $J_P$ is not cotangent to $P$, because $P$ is not a subgroup, so that it will have a component
tangent to the fiber $H$. Fortunately, our choice for the product separates 
 $P$ from $H$ orthogonally, and we 
can obtain the infinitesimal displacement along $P$ by an orthogonal projection (that is dropping the terms
$T_i$, $i=1,2,3$). If we call  such a projection $\Pi$, we, thus, have
$$
ds^2_P=-\frac 12 {\rm Trace}(\Pi (J_P) \Pi (J_P)),
$$
where
$$
J_P=dx_3 T_4-\sin x_3 dx_2 T_6 -\sin x_2 \sin x_3 T_5 dx_1 +\ldots,
$$
the dots indicating the terms tangent to $H$. Thus, we get
$$
ds^2_P =dx_3^2 +\sin^2 x_3 (dx_2^2 +\sin^2 x_2 dx_1^2),
$$
which is just the metric of $S^3$ in the usual spherical polar coordinates,
$$
S^3=\{X\in \mathbb {R}^4| X_1=\cos x_3 ,\ X_2 =\sin x_3 \cos x_2,\ X_3 =\sin x_3 \sin x_2 \cos x_1,\ X_4 =\sin x_3 \sin x_2 \sin x_1 \},
$$
so that we must choose
$$
x_3\in [0,\pi], \qquad x_2\in [0,\pi], \qquad x_3\in [0,2\pi].
$$
This complete our determination of the ranges. \\

The invariant measure is
\begin{equation} \label{invariant}
d\mu_{SO(4)}=\sqrt {{\rm det} ds^2_H}\sqrt {{\rm det} ds^2_P}=\sin x_2 
\sin^2 x_3 \sin x_5 dx_1 dx_2 dx_3 dx_4 dx_5 dx_6.
\end{equation}
Note that there is a second quite interesting way to determine the ranges using the measure. To understand it, let us think
of a sphere $S^2$ parametrised with latitudinal and longitudinal (polar) coordinates. But suppose we do not know the ranges
of parameters. The only fact we know is that we can cover  our sphere entirely, if any coordinate runs over a whole period
($\lambda\in [0,2\pi]$, $\phi\in [0,2\pi]$). The problem is that we can cover the sphere many times (twice in our case).
But suppose we know the corresponding measure: $|\sin \lambda| d\lambda d\phi$. We see 
that it becomes 
singular at $\lambda=0$ 
and $\lambda=\pi$. These 
correspond to the points 
where the parallels shrink down (that is, the north and south poles). This implies
that when $\lambda$ runs over $[0,\pi]$, we generate a closed surface which, thus, must cover the sphere (which contains no
closed surfaces but itself), so that the range of $\lambda$ can be restricted to $[0,\pi]$.\\
In the same way, we can look at the invariant metric we have just constructed for $SO(4)$. All coordinates $x_I$, $I=1,\ldots,6$
have period $[0,2\pi]$ because 
the $\exp (x_I T_I)$ 
are functions of $\cos x_I$ and $\sin x_I$. However, $x_2$, $x_3$ and
$x_5$ appear in the measure exactly as for the sphere and must be restricted to $[0,\pi]$. Thus, we re-obtain the ranges
\begin{equation} \label{ranges}
x_1,x_4,x_6\in[0,2\pi],\qquad x_2,x_3,x_5\in[0,\pi].
\end{equation}

\section{Appendix II (S. Cacciatori)}
Let us call $\Delta$ the eigenvalues simplex and $M$ the subset of varying angles for the chosen example. Finally let us call $S$ the subset
of $M$ imposed by the Peres-Horodecki condition (\ref{PH2})
\begin{equation} 
-\frac{1}{4} \sin ^2\left(2 x_5\right) \left(\lambda _1-\lambda
   _3\right){}^2-\lambda _2 \left(\lambda _1+\lambda _2+\lambda
   _3-1\right) \geq 0.
\end{equation}
Thus we need to compute
\begin{equation}
ESF_{\{x_5\}}=\int_S d\mu/\int_M d\mu
\end{equation}
$d\mu$ being the measure on the given region. In particular (\ref{PH2}) constraints the range of $x_5$ only, so that
\begin{equation}
ESF_{\{x_5\}}=\frac 12 \int_S \sin x_5 dx_5.
\end{equation}
Note that (\ref{PH2}) is invariant under $x_5\rightarrow \pi-x_5$ so that we can restrict the region $S $ to $S_0=S\cap [0,\pi/2]$ and write
\begin{equation}
ESF_{\{x_5\}}=\int_{S_0} \sin x_5 dx_5.
\end{equation}
Furthermore, in $[0,\pi/2]$ the map $y_5=\sin x_5$ is bijective as a map $[0,\pi/2]\rightarrow[0,1]$ so that we can write
\begin{equation}
ESF_{\{x_5\}}=\int_{\sigma} dy_5,
\end{equation}
where $\sigma$ is the set of solutions of
\begin{equation} \label{sigma}
-y_5^2 (1-y_5^2) \left(\lambda _1-\lambda_3\right){}^2+\lambda _2 \lambda_4 \geq 0,
\end{equation}
in $[0,1]$, $\lambda_4=1-\lambda_1-\lambda_2-\lambda_3$. To determine $\sigma$ let us first set $z=y_5^2$.
The solutions of
\begin{equation}
(z^2-z) \left(\lambda _1-\lambda_3\right){}^2+\lambda _2 \lambda_4 = 0,
\end{equation}
are
\begin{equation}
z_\pm =\frac 12 \left[1\pm \sqrt{1-\frac {4\lambda_2 \lambda_4}{(\lambda_1-\lambda_3)^2}} \right].
\end{equation}
When the discriminant is non positive, that is when $(\lambda_1-\lambda_3)^2\leq 4\lambda_2\lambda_4$, then (\ref{sigma}) is satisfied for all
$y$ and $ESF=1$. Otherwise $z_\pm$ satisfy
\begin{equation}
0\leq z_- < z_+ \leq 1.
\end{equation}
In this case
\begin{equation}
(z^2-z) \left(\lambda _1-\lambda_3\right){}^2+\lambda _2 \lambda_4 \geq 0,
\end{equation}
is solved for external values and being $z\in[0,1]$ we have that (\ref{sigma}) is satisfied for
\begin{equation}
z\in [0,z_-]\cup [z_+,1],
\end{equation}
and then (being again $\sqrt {z}=y_5\in [0,1]$)
\begin{equation}
y_5\in [0,\sqrt {z_-}]\cup [\sqrt{z_+},1]\equiv \sigma.
\end{equation}
Thus
\begin{equation}
ESF_{\{x_5\}}=1-\sqrt{z_+}+\sqrt {z_-}=1-\frac{\sqrt{1+\sqrt{1-U ^2}}}{\sqrt{2}}+\frac{\sqrt{1-\sqrt{1-U^2}}}{\sqrt{2}}.
\end{equation}

\begin{acknowledgments}
I would like to express appreciation to the Kavli Institute for Theoretical
Physics (KITP)
for computational support in this research, and to Sergio Cacciatori for deriving and communicating to me his Euler-angle parameterization of $SO(4)$, and 
permitting it to be presented in Appendix I. 
Also, K. {\.Z}yczkowski alerted me to the importance of the 
absolute separability bound of Verstraete, Audenaert and De Moor \cite{ver}, 
while M. Trott was remarkably supportive in assisting with Mathematica
computations. Also, Vladeta Jovovic was very helpful in manipulating
trigonometric expressions.
\end{acknowledgments}

\bibliography{Param4}

\begin{thebibliography}{37}
\expandafter\ifx\csname natexlab\endcsname\relax\def\natexlab#1{#1}\fi
\expandafter\ifx\csname bibnamefont\endcsname\relax
  \def\bibnamefont#1{#1}\fi
\expandafter\ifx\csname bibfnamefont\endcsname\relax
  \def\bibfnamefont#1{#1}\fi
\expandafter\ifx\csname citenamefont\endcsname\relax
  \def\citenamefont#1{#1}\fi
\expandafter\ifx\csname url\endcsname\relax
  \def\url#1{\texttt{#1}}\fi
\expandafter\ifx\csname urlprefix\endcsname\relax\def\urlprefix{URL }\fi
\providecommand{\bibinfo}[2]{#2}
\providecommand{\eprint}[2][]{\url{#2}}

\bibitem[{\citenamefont{{\.Z}yczkowski
  et~al.}(1998)\citenamefont{{\.Z}yczkowski, Horodecki, Sanpera, and
  Lewenstein}}]{ZHSL}
\bibinfo{author}{\bibfnamefont{K.}~\bibnamefont{{\.Z}yczkowski}},
  \bibinfo{author}{\bibfnamefont{P.}~\bibnamefont{Horodecki}},
  \bibinfo{author}{\bibfnamefont{A.}~\bibnamefont{Sanpera}}, \bibnamefont{and}
  \bibinfo{author}{\bibfnamefont{M.}~\bibnamefont{Lewenstein}},
  \bibinfo{journal}{Phys. Rev. A} \textbf{\bibinfo{volume}{58}},
  \bibinfo{pages}{883} (\bibinfo{year}{1998}).

\bibitem[{\citenamefont{Slater}(1999{\natexlab{a}})}]{slaterHall}
\bibinfo{author}{\bibfnamefont{P.~B.} \bibnamefont{Slater}},
  \bibinfo{journal}{J. Phys. A} \textbf{\bibinfo{volume}{32}},
  \bibinfo{pages}{8231} (\bibinfo{year}{1999}{\natexlab{a}}).

\bibitem[{\citenamefont{Slater}(1999{\natexlab{b}})}]{slaterA}
\bibinfo{author}{\bibfnamefont{P.~B.} \bibnamefont{Slater}},
  \bibinfo{journal}{J. Phys. A} \textbf{\bibinfo{volume}{32}},
  \bibinfo{pages}{5261} (\bibinfo{year}{1999}{\natexlab{b}}).

\bibitem[{\citenamefont{Slater}(2000{\natexlab{a}})}]{slaterC}
\bibinfo{author}{\bibfnamefont{P.~B.} \bibnamefont{Slater}},
  \bibinfo{journal}{Euro. Phys. J. B} \textbf{\bibinfo{volume}{17}},
  \bibinfo{pages}{471} (\bibinfo{year}{2000}{\natexlab{a}}).

\bibitem[{\citenamefont{Slater}(2000{\natexlab{b}})}]{slaterOptics}
\bibinfo{author}{\bibfnamefont{P.~B.} \bibnamefont{Slater}},
  \bibinfo{journal}{J. Opt. B} \textbf{\bibinfo{volume}{2}},
  \bibinfo{pages}{L19} (\bibinfo{year}{2000}{\natexlab{b}}).

\bibitem[{\citenamefont{Slater}(2005{\natexlab{a}})}]{slaterJGP}
\bibinfo{author}{\bibfnamefont{P.~B.} \bibnamefont{Slater}},
  \bibinfo{journal}{J. Geom. Phys.} \textbf{\bibinfo{volume}{53}},
  \bibinfo{pages}{74} (\bibinfo{year}{2005}{\natexlab{a}}).

\bibitem[{\citenamefont{Slater}(2005{\natexlab{b}})}]{slaterPRA}
\bibinfo{author}{\bibfnamefont{P.~B.} \bibnamefont{Slater}},
  \bibinfo{journal}{Phys. Rev. A} \textbf{\bibinfo{volume}{71}},
  \bibinfo{pages}{052319} (\bibinfo{year}{2005}{\natexlab{b}}).

\bibitem[{\citenamefont{Slater}(2006)}]{pbsCanosa}
\bibinfo{author}{\bibfnamefont{P.~B.} \bibnamefont{Slater}},
  \bibinfo{journal}{J. Phys. A} \textbf{\bibinfo{volume}{39}},
  \bibinfo{pages}{913} (\bibinfo{year}{2006}).

\bibitem[{\citenamefont{Slater}(2007{\natexlab{a}})}]{slaterPRA2}
\bibinfo{author}{\bibfnamefont{P.~B.} \bibnamefont{Slater}},
  \bibinfo{journal}{Phys. Rev. A} \textbf{\bibinfo{volume}{75}},
  \bibinfo{pages}{032326} (\bibinfo{year}{2007}{\natexlab{a}}).

\bibitem[{\citenamefont{Slater}(2007{\natexlab{b}})}]{slater833}
\bibinfo{author}{\bibfnamefont{P.~B.} \bibnamefont{Slater}},
  \bibinfo{journal}{J. Phys. A} \textbf{\bibinfo{volume}{40}},
  \bibinfo{pages}{14279} (\bibinfo{year}{2007}{\natexlab{b}}).

\bibitem[{\citenamefont{Slater}(2008)}]{slaterJGP2}
\bibinfo{author}{\bibfnamefont{P.~B.} \bibnamefont{Slater}},
  \bibinfo{journal}{J. Geom. Phys.}
  \textbf{\bibinfo{volume}{doi:10.1016/j.geomphys.2008.03.014}},
  \bibinfo{pages}{1} (\bibinfo{year}{2008}).

\bibitem[{\citenamefont{Wootters}(1990)}]{wkw1}
\bibinfo{author}{\bibfnamefont{W.~K.} \bibnamefont{Wootters}},
  \bibinfo{journal}{Found. Phys.} \textbf{\bibinfo{volume}{20}},
  \bibinfo{pages}{1365} (\bibinfo{year}{1990}).

\bibitem[{\citenamefont{Kass}(1989)}]{kass}
\bibinfo{author}{\bibfnamefont{R.~E.} \bibnamefont{Kass}},
  \bibinfo{journal}{Statist. Sci.} \textbf{\bibinfo{volume}{4}},
  \bibinfo{pages}{188} (\bibinfo{year}{1989}).

\bibitem[{\citenamefont{Kwek et~al.}(1999)\citenamefont{Kwek, Oh, and
  Wang}}]{kwek}
\bibinfo{author}{\bibfnamefont{L.~C.} \bibnamefont{Kwek}},
  \bibinfo{author}{\bibfnamefont{C.~H.} \bibnamefont{Oh}}, \bibnamefont{and}
  \bibinfo{author}{\bibfnamefont{X.-B.} \bibnamefont{Wang}},
  \bibinfo{journal}{J. Phys. A} \textbf{\bibinfo{volume}{32}},
  \bibinfo{pages}{6613} (\bibinfo{year}{1999}).

\bibitem[{\citenamefont{Ozawa}(2000)}]{ozawa}
\bibinfo{author}{\bibfnamefont{M.}~\bibnamefont{Ozawa}},
  \bibinfo{journal}{Phys. Lett. A} \textbf{\bibinfo{volume}{268}},
  \bibinfo{pages}{158} (\bibinfo{year}{2000}).

\bibitem[{\citenamefont{{\.Z}yczkowski and Sommers}(2003)}]{szHS}
\bibinfo{author}{\bibfnamefont{K.}~\bibnamefont{{\.Z}yczkowski}}
  \bibnamefont{and} \bibinfo{author}{\bibfnamefont{H.-J.}
  \bibnamefont{Sommers}}, \bibinfo{journal}{J. Phys. A}
  \textbf{\bibinfo{volume}{36}}, \bibinfo{pages}{10115} (\bibinfo{year}{2003}).

\bibitem[{\citenamefont{Petz and Sud{\'a}r}(1996)}]{petzsudar}
\bibinfo{author}{\bibfnamefont{D.}~\bibnamefont{Petz}} \bibnamefont{and}
  \bibinfo{author}{\bibfnamefont{C.}~\bibnamefont{Sud{\'a}r}},
  \bibinfo{journal}{J. Math. Phys.} \textbf{\bibinfo{volume}{37}},
  \bibinfo{pages}{2662} (\bibinfo{year}{1996}).

\bibitem[{\citenamefont{Sommers and {\.Z}yczkowski}(2003)}]{szBures}
\bibinfo{author}{\bibfnamefont{H.-J.} \bibnamefont{Sommers}} \bibnamefont{and}
  \bibinfo{author}{\bibfnamefont{K.}~\bibnamefont{{\.Z}yczkowski}},
  \bibinfo{journal}{J. Phys. A} \textbf{\bibinfo{volume}{36}},
  \bibinfo{pages}{10083} (\bibinfo{year}{2003}).

\bibitem[{\citenamefont{Bengtsson and {\.Z}yczkowski}(2006)}]{ingemarkarol}
\bibinfo{author}{\bibfnamefont{I.}~\bibnamefont{Bengtsson}} \bibnamefont{and}
  \bibinfo{author}{\bibfnamefont{K.}~\bibnamefont{{\.Z}yczkowski}},
  \emph{\bibinfo{title}{Geometry of Quantum States}}
  (\bibinfo{publisher}{Cambridge}, \bibinfo{address}{Cambridge},
  \bibinfo{year}{2006}).

\bibitem[{\citenamefont{Horn and Johnson}(1991)}]{hornjohnson}
\bibinfo{author}{\bibfnamefont{R.~A.} \bibnamefont{Horn}} \bibnamefont{and}
  \bibinfo{author}{\bibfnamefont{C.~R.} \bibnamefont{Johnson}},
  \emph{\bibinfo{title}{Matrix Analysis}} (\bibinfo{publisher}{Cambridge
  Univ.}, \bibinfo{address}{New York}, \bibinfo{year}{1991}).

\bibitem[{\citenamefont{Nielsen and Vidal}(2001)}]{nielsenvidal}
\bibinfo{author}{\bibfnamefont{M.~A.} \bibnamefont{Nielsen}} \bibnamefont{and}
  \bibinfo{author}{\bibfnamefont{G.}~\bibnamefont{Vidal}},
  \bibinfo{journal}{Quant. Inform. Comput.} \textbf{\bibinfo{volume}{1}},
  \bibinfo{pages}{76} (\bibinfo{year}{2001}).

\bibitem[{\citenamefont{Tilma et~al.}(2002)\citenamefont{Tilma, Byrd, and
  Sudarshan}}]{tbs}
\bibinfo{author}{\bibfnamefont{T.}~\bibnamefont{Tilma}},
  \bibinfo{author}{\bibfnamefont{M.}~\bibnamefont{Byrd}}, \bibnamefont{and}
  \bibinfo{author}{\bibfnamefont{E.~C.~G.} \bibnamefont{Sudarshan}},
  \bibinfo{journal}{J. Phys. A} \textbf{\bibinfo{volume}{35}},
  \bibinfo{pages}{10445} (\bibinfo{year}{2002}).

\bibitem[{\citenamefont{Batle et~al.}(2002)\citenamefont{Batle, Plastino,
  Casas, and Plastino}}]{batleplastino1}
\bibinfo{author}{\bibfnamefont{J.}~\bibnamefont{Batle}},
  \bibinfo{author}{\bibfnamefont{A.~R.} \bibnamefont{Plastino}},
  \bibinfo{author}{\bibfnamefont{M.}~\bibnamefont{Casas}}, \bibnamefont{and}
  \bibinfo{author}{\bibfnamefont{A.}~\bibnamefont{Plastino}},
  \bibinfo{journal}{Phys. Lett. A} \textbf{\bibinfo{volume}{298}},
  \bibinfo{pages}{301} (\bibinfo{year}{2002}).

\bibitem[{\citenamefont{Peres}(1996)}]{asher}
\bibinfo{author}{\bibfnamefont{A.}~\bibnamefont{Peres}},
  \bibinfo{journal}{Phys. Rev. Lett.} \textbf{\bibinfo{volume}{77}},
  \bibinfo{pages}{1413} (\bibinfo{year}{1996}).

\bibitem[{\citenamefont{Horodecki et~al.}(1996)\citenamefont{Horodecki,
  Horodecki, and Horodecki}}]{michal}
\bibinfo{author}{\bibfnamefont{M.}~\bibnamefont{Horodecki}},
  \bibinfo{author}{\bibfnamefont{P.}~\bibnamefont{Horodecki}},
  \bibnamefont{and}
  \bibinfo{author}{\bibfnamefont{R.}~\bibnamefont{Horodecki}},
  \bibinfo{journal}{Phys. Lett. A} \textbf{\bibinfo{volume}{223}},
  \bibinfo{pages}{1} (\bibinfo{year}{1996}).

\bibitem[{\citenamefont{Ishizaka and Hiroshima}(2000)}]{ishi}
\bibinfo{author}{\bibfnamefont{S.}~\bibnamefont{Ishizaka}} \bibnamefont{and}
  \bibinfo{author}{\bibfnamefont{T.}~\bibnamefont{Hiroshima}},
  \bibinfo{journal}{Phys. Rev. A} \textbf{\bibinfo{volume}{62}},
  \bibinfo{pages}{022310} (\bibinfo{year}{2000}).

\bibitem[{\citenamefont{Verstraete et~al.}(2001)\citenamefont{Verstraete,
  Audenaert, and Moor}}]{ver}
\bibinfo{author}{\bibfnamefont{F.}~\bibnamefont{Verstraete}},
  \bibinfo{author}{\bibfnamefont{K.}~\bibnamefont{Audenaert}},
  \bibnamefont{and} \bibinfo{author}{\bibfnamefont{B.~D.} \bibnamefont{Moor}},
  \bibinfo{journal}{Phys. Rev. A} \textbf{\bibinfo{volume}{64}},
  \bibinfo{pages}{012316} (\bibinfo{year}{2001}).

\bibitem[{\citenamefont{Hildebrand}(2007)}]{roland}
\bibinfo{author}{\bibfnamefont{R.}~\bibnamefont{Hildebrand}},
  \bibinfo{journal}{Phys. Rev. A} \textbf{\bibinfo{volume}{76}},
  \bibinfo{pages}{052325} (\bibinfo{year}{2007}).

\bibitem[{\citenamefont{Batle et~al.}(2004)\citenamefont{Batle, Plastino,
  Casas, and Plastino}}]{inclusion}
\bibinfo{author}{\bibfnamefont{J.}~\bibnamefont{Batle}},
  \bibinfo{author}{\bibfnamefont{A.~R.} \bibnamefont{Plastino}},
  \bibinfo{author}{\bibfnamefont{M.}~\bibnamefont{Casas}}, \bibnamefont{and}
  \bibinfo{author}{\bibfnamefont{A.}~\bibnamefont{Plastino}},
  \bibinfo{journal}{J. Phys. A} \textbf{\bibinfo{volume}{37}},
  \bibinfo{pages}{895} (\bibinfo{year}{2004}).

\bibitem[{\citenamefont{Cl{\'e}ment and Raggio}(2006)}]{clement}
\bibinfo{author}{\bibfnamefont{M.~E.~G.} \bibnamefont{Cl{\'e}ment}}
  \bibnamefont{and} \bibinfo{author}{\bibfnamefont{G.~A.}
  \bibnamefont{Raggio}}, \bibinfo{journal}{J. Phys. A}
  \textbf{\bibinfo{volume}{39}}, \bibinfo{pages}{9291} (\bibinfo{year}{2006}).

\bibitem[{\citenamefont{Raggio}(2006)}]{raggio}
\bibinfo{author}{\bibfnamefont{G.~A.} \bibnamefont{Raggio}},
  \bibinfo{journal}{J. Phys. A} \textbf{\bibinfo{volume}{39}},
  \bibinfo{pages}{617} (\bibinfo{year}{2006}).

\bibitem[{\citenamefont{Headrick and Wiseman}(2005)}]{headrick}
\bibinfo{author}{\bibfnamefont{M.}~\bibnamefont{Headrick}} \bibnamefont{and}
  \bibinfo{author}{\bibfnamefont{T.}~\bibnamefont{Wiseman}},
  \bibinfo{journal}{Class. Quant. Grav.} \textbf{\bibinfo{volume}{22}},
  \bibinfo{pages}{4931} (\bibinfo{year}{2005}).

\bibitem[{\citenamefont{Szarek et~al.}(2006)\citenamefont{Szarek, Bengtsson,
  and {\.Z}yczkowski}}]{sbz}
\bibinfo{author}{\bibfnamefont{S.}~\bibnamefont{Szarek}},
  \bibinfo{author}{\bibfnamefont{I.}~\bibnamefont{Bengtsson}},
  \bibnamefont{and}
  \bibinfo{author}{\bibfnamefont{K.}~\bibnamefont{{\.Z}yczkowski}},
  \bibinfo{journal}{J. Phys. A} \textbf{\bibinfo{volume}{39}},
  \bibinfo{pages}{L119} (\bibinfo{year}{2006}).

\bibitem[{\citenamefont{Innami}(1999)}]{innami}
\bibinfo{author}{\bibfnamefont{N.}~\bibnamefont{Innami}},
  \bibinfo{journal}{Proc. Amer. Math. Soc.} \textbf{\bibinfo{volume}{127}},
  \bibinfo{pages}{3049} (\bibinfo{year}{1999}).

\bibitem[{\citenamefont{Batle et~al.}(2006)\citenamefont{Batle, Casas,
  Plastino, and Plastino}}]{batlecasas1}
\bibinfo{author}{\bibfnamefont{J.}~\bibnamefont{Batle}},
  \bibinfo{author}{\bibfnamefont{M.}~\bibnamefont{Casas}},
  \bibinfo{author}{\bibfnamefont{A.}~\bibnamefont{Plastino}}, \bibnamefont{and}
  \bibinfo{author}{\bibfnamefont{A.~R.} \bibnamefont{Plastino}},
  \bibinfo{journal}{Phys. Lett. A} \textbf{\bibinfo{volume}{353}},
  \bibinfo{pages}{161} (\bibinfo{year}{2006}).

\bibitem[{\citenamefont{Dahl et~al.}(2007)\citenamefont{Dahl, Leinaas, Myrheim,
  and Ovrum}}]{dahl}
\bibinfo{author}{\bibfnamefont{G.}~\bibnamefont{Dahl}},
  \bibinfo{author}{\bibfnamefont{J.~M.} \bibnamefont{Leinaas}},
  \bibinfo{author}{\bibfnamefont{J.}~\bibnamefont{Myrheim}}, \bibnamefont{and}
  \bibinfo{author}{\bibfnamefont{E.}~\bibnamefont{Ovrum}},
  \bibinfo{journal}{Lin. Alg. Applic.} \textbf{\bibinfo{volume}{420}},
  \bibinfo{pages}{711} (\bibinfo{year}{2007}).

\bibitem[{\citenamefont{Bernardoni et~al.}()\citenamefont{Bernardoni,
  Cacciatori, Cerchiai, and Scotti}}]{sergioF4}
\bibinfo{author}{\bibfnamefont{F.}~\bibnamefont{Bernardoni}},
  \bibinfo{author}{\bibfnamefont{S.~L.} \bibnamefont{Cacciatori}},
  \bibinfo{author}{\bibfnamefont{B.~L.} \bibnamefont{Cerchiai}},
  \bibnamefont{and} \bibinfo{author}{\bibfnamefont{A.}~\bibnamefont{Scotti}},
  \eprint{arXiv:0705.3978v2}.

\end{thebibliography}

\end{document}